\documentclass{basi} \usepackage{txfonts} \let\new=\newcommand
\new{\be}{\begin{equation} \displaystyle} \new{\ee}{\end{equation}}
\new{\bea}{\begin{eqnarray}} \new{\eea}{\end{eqnarray}}
\new{\bfig}{\begin{figure}} \new{\efig}{\end{figure}}
\new{\lsim}{\raisebox{-0.3ex}{\mbox{$\stackrel{<}{_\sim} \,$}}}
\new{\gsim}{\raisebox{-0.3ex}{\mbox{$\stackrel{>}{_\sim} \,$}}}
\new{\del}{\partial} \new{\diff}{{\rm d}} \new{\dt}{\rm{d}_t}
\new{\cM}{{\cal M}} \new{\bhs}{{\texttt{\large m}}_{h8}}
\new{\bh}{{\texttt{\large m}}_{h}} \new{\Msun}{M_\odot}
\new{\dif}{{\rm d}x^3}
\begin{document}
\title[Supermassive Disks]{Physics of Supermassive Disks: Formation
and Collapse}  
\author[A. Mangalam]
{Arun Mangalam \thanks{e-mail:mangalam@iiap.ernet.in} \\  Indian Institute of
Astrophysics, Koramangala, Bangalore 560 034}
\maketitle
\label{firstpage}
\begin{abstract}
Supermassive disks are thought to be precursors of supermassive black
holes that are believed to power quasars and exist at centers of
galaxies. Formation scenarios of such disks are reviewed and it is
argued that gas dynamical  schemes are favourable compared to stellar
dynamical schemes which could however be important feeding mechanisms
for the growth of the black hole.  A new self-similar model of a
collapse of a self-gravitating disk due to radiation induced  stresses
applicable to two different situations of radiative viscosity and
Compton drag  is presented. The collapse timescale purely due to
radiative viscosity is found to be a  fraction of Hubble time,
$\tau_\gamma \sim{\sigma_T c/ (m_p G)} ({L_{edd}/ L}) \simeq 6 \times
10^9$yrs is slow and probably magnetic fields play an
important role before general relativistic effects take over.  A model
of self-gravitating disk collapsing due to Compton drag by the Cosmic
Microwave Background is also presented which is found to be effective
at redshifts $1400>z\gsim 300$. It is proposed that the small  $\lsim
10^5 \Msun$ objects that form by this mechanism by $z \sim 20$ can
merge and  coalesce by dynamical friction to form the high redshift
quasars seen. Supermassive stars which are systems (and could be end
products of a supermassive disk phase) en route to the  final collapse
are also briefly reviewed.

\end{abstract}

\section{Introduction}
\label{sec:intro}
There seems to be increasing evidence that supermassive black holes
   are at the  centers of galaxies. Dynamical searches indicate the
   existence of massive dark  objects (MDOs) in eight systems and
   their masses range from $10^6-10^{9.5} \Msun$  (Kormendy \&
   Richstone 1995). Although this study does not confirm that the
   central  objects are supermassive black holes, it has been inferred
   that the central mass is contained within  $10^5$ Schwarzchild
   radii.   On an  average, the black hole mass is a fraction,
   $10^{-2}-10^{-3}$, of the total mass of the  galaxy and of order
   $10^{-3.5}$ of the bulge mass (Wandel 1999). Recent observations
   show a strong correlation between the black hole mass, $\bh$, from
   stellar dynamical estimates, and the velocity dispersion of the
   host bulges ($\bh \propto \sigma^\alpha$; where $\alpha$ is
   reported to be in the range 3.5--5; eg. Ferrarese \& Merritt, 2000).

 The presence of quasars at high redshifts tells us that galaxy
formation had proceeded far enough for supermassive black holes to
form in the standard picture (Rees 1984). A detailed model of
formation of these objects should address the issues of supernovae
feedback from star formation and the mechanism of efficient angular
momentum transport in order to explain the massive active nuclei as
early as $z=6$. In the case of MDOs, there is a need to explain the
compact sizes of $10-100$ pc that are implied from dynamical studies.

\section{Formation scenarios}
\label{schemes}
We first discuss the possibility that dense star clusters feed a seed
black hole which grows initially by accreting the gas resulting from
tidal disruption that occurs at the tidal radius, $R_T \simeq (6 \bh/
\pi\rho_*$). Subsequently, when the Schwarzchild radius,$R_s$, exceeds
this, the black hole then grows by swallowing low angular momentum
stars whose pericenters   lie within $R_s$. The timescale for growth
during the gas release through tidal disruption suffers from serious
drawbacks-- the accretion rate is given by (see eg. Hills 1975),
$\dot{M} = (4 \pi^2 G \bh R_T/\sigma) \rho_s$,  assuming a Maxwellian
distribution of velocities with dispersion $\sigma$.  At this rate,
the growth timescale goes as $~1.7\times 10^{11} (\rho_s/10^6  \Msun
{\rm pc}^{-3})(\bh/\Msun)^{1/3}(\sigma/{\rm 100 km/sec})$yrs which is
large even for a dense cluster (the alternative of a seed mass of 1000
$\Msun$ for the black hole is unlikely from theories of stellar
evolution). Such models then beg  the question of how clusters of
densities $\sim 10^7 \Msun/{\rm pc}^3$ form in  the first place.

Now consider the situation where the growth proceeds by swallowing of
stars.  Swallowing rate of stars may be estimated in the following
way. Stellar orbits diffuse by two-body relaxation toward lower
angular momentum orbits until they enter a small loss-cone of
semi-aperture $\theta_c \simeq (t_{dyn}/t_R)^{1/2}$ (Frank \& Rees
1976, Lightman \& Shapiro 1977) where $t_R$ is the relaxation time and
$t_{dyn}$ is the dynamical time. The resulting swallowing rate is
$\dot{M} \approx N_c m_*/( t_R \ln{(2/\theta_c)})\approx m_*/t_{dyn}$
which is not rapid enough in  most cases. The key point is that in
this scenario, one must assume a  extremely dense and massive cluster.

Recently, there have been proposals (Volonteri, Haardt and Madau 2003,
Wyithe and Loeb 2002) motivated by quasars discovered at $z \approx 6$
that these objects have are assemblies of smaller $~100 \Msun$ objects
that collapsed at $z \sim 20$ from  high-$\sigma$ density
fluctuations. The former model invokes dynamical friction of merging
systems to sink to the center as larger halo objects and involves
Monte-Carlo computations based on halo merger trees from cosmological
simulations  in a $\Lambda$ CDM cosmology. The model appears to yield
the desired results of  luminosity functions. This promising model for
forming quasars at  high red-shifts is worth exploring further to
ascertain how massive the seed black holes need to be to explain the
high redshift quasars and the efficiency of mergers. Observations of
ultra luminous X-ray sources (ULXs) by (Colbert \& Mushotsky 1999,
Makishima et al. 2000) in nearby galaxies  seem to indicate that seed
black holes of intermediate mass of a few hundred $\Msun$ are possible.

There are good reasons to think that supermassive gaseous objects 
are remnants of a galaxy formation process. Mangalam(2001) presented
a detailed physical model wherein protoquasars (or MDOs) form from a
magnetized  accretion of a collapsed disk, the properties of which are
obtained taking into account supernovae feedback in a virialized
halo. Significant star formation and supernovae activity occurs after
the cloud, which is spun up by tidal torques, contracts to a radius
where self-gravity is significant. The model is composed of the
following stages-

\begin{enumerate}
\item 
\noindent
The formation of a gaseous disk with a radial extent of about a kpc,
 in a host galaxy as limited by supernovae feed back. The range in
 halo mass for a given redshift that still retains the hot gas was
 calculated.
\item
\noindent
In previous work, gravitational instabilities  in the disk was
considered as the main source of viscosity.  Justification was made
for a magnetic viscosity and the estimated accretion rate turns out to
be significant. The collapse of the disk was calculated with a
generalized viscosity prescription (which includes the individual
cases of magnetic, $\alpha$ and self-gravity induced instabilities,
under a halo dominated gravitational potential into a compact central
region at rapid rate of about a $\Msun~{\rm yr}^{-1}$. A
self-gravitating magnetized disk solution for this central object that
collapses to a seed black hole in $10^6$ yrs, was calculated.

\item
\noindent
The implications for quasar luminosity functions and the time delay
between collapse and virialization is considered in Mangalam(2003) and
is based on the mass limits from cooling considerations in
Mangalam(2001).
\end{enumerate}

\section{Collapse of supermassive disks}
\label{collapse}
Here we calculate the collapse of self-gravitating compact mass that
takes into account radiative stresses, which is a  Newtonian 1.5
dimensional version of a quasi-spherical relativistic collapse
currently under investigation. A particular application can be made to
disks collapsing under angular momentum transport by radiative drag
due to CMBR at high redshift. Another application is to estimate
collapse time scale due to radiative viscosity after sufficient
accretion of mass into a compact region of radius $r_0$, typically of
the order of a hundred parsecs containing a mass of $10^8 \Msun$. The
problem of self-gravitating accretion flow is complicated by the
coupling of Poisson's equation to the momentum and continuity
equations. Clearly, its evolution has to be treated differently from
the case of a prescribed background potential.

   We consider a simpler model of disk where the self-consistent
density distribution with a gravitational potential that is entirely
due to self-gravity, is of the Mestel form $\Sigma(r,t)= {v_\phi^2(t)/
(2\pi G r)}$, where the time dependence appears only in the rotational
velocity.  Taking $v_\phi= v_0 \,\chi(t)$ and $r=r_0 \, \chi_1(t) \,
x$ where $v_0^2= G M_c/r_0$, and $M_c$ is the mass out to $r_0$. We
see that by assuming a self-similar evolution of the disk, the mass
out to a given $x$ should be independent of $t$ and hence it follows
that $\chi_1 = \chi^{-2}$ and $\Sigma= \Sigma_c \,{\chi^4/ x}$ where
$\Sigma_c= M_c/ (2\pi r_0^2)$.  From the continuity equation,
\begin{equation}
	       r\del_t \Sigma + \del_r(r \Sigma v_r)=0
\end{equation} 
			       we find
\begin{equation}
		 v_r= -4 \chi^{-3} \dot{\chi} r_0 x.
\end{equation}
  Substituting this and the self-similar forms given above into the
		      angular momentum equation,
\begin{equation}
\Sigma \del_t v_\phi +\Sigma (v_r/r) \del_r(r v_\phi)=
(1/r^2)\del_r(r^2  \Pi_{r\phi})
\end{equation}
			      we obtain
\begin{equation}
\Pi_{r\phi}=-{3 \over 2} v_0 \Sigma_c r_0 \chi^2 \dot{\chi},
\label{chi}
\end{equation}
   which is {\em independent} of $x$. So far no specific viscosity
   mechanism has been invoked-- the form of $\Pi_{r\phi}$ above is
   necessitated by the prescription of a Mestel disk. If a  stress due
   radiative viscosity is assumed, $\Pi_{r\phi}= \eta_\gamma r H
   {\diff \Omega \over \diff r}$,  where the coefficient of radiative
   viscosity (Misner 1968, Weinberg 1971) $\eta_\gamma=(8/27) (
   \epsilon_\gamma)/(\sigma_T n_e c)$, where  $\epsilon_\gamma$ is the
   photon energy density, $H$ is the half thickness, $\sigma_T$ is
   the Thomson cross-section and $n_e$ is the electron density. From the energy dissipation  condition,
   the heat flux is given by
\be \nabla \cdot {\bf F} \simeq \eta_\gamma \left(
   \frac{v_\phi}{r} \right)^2, 
\ee  taking only the relevant
   component of the heat flux, \be {\bf F}=-\frac{c}{n_e \sigma_T}
   \nabla p_\gamma \ee for Thomson scattering opacity.   It follows
   that the half thickness, $H \propto x  \chi(t)^{-3}$. Realistically
   the 1.5-D assumption breaks down. Nevertheless, we can push our
   model to get  some estimates. It follows that $\eta_\gamma \propto
   \epsilon_\gamma/\rho c^2 \propto x$. Since it is radiation
   dominated, we assume a polytrope of index 4/3 and obtain 
\be
   \frac{2}{3} \tau_\gamma ~\chi^2~ \dot{\chi}= \chi^{7/3}, 
\ee which
   leads to the solution \be \chi(t)=  \left ( {t \over \tau_\gamma}+1
   \right)^{3/2}, \ee where $\chi(0)=1$ was taken as the initial
   condition. The factor $\epsilon_\gamma/(\rho c^2)$ can be estimated by
calculating the luminosity due to the heat flux by 
taking $\epsilon_\gamma=3 p_\gamma$.
The collapse timescale is then given by 
\be
   \tau_\gamma \simeq{\sigma_T c \over m_p G} {2 \over 3 \pi}{L_{edd} \over L}
{r_g \over r_0} \sim 6 \times 10^{9} {\rm yrs} 
\ee 
where $r_g = G M_c/c^2$, the gravitational radius; the fiducial value 
taken here corresponds to a situation when the system is 
sufficiently compact and radiating
at a tenth of eddington luminosity.  Clearly, the model is
   not strictly  valid when it is relativistic.  The toy Newtonian
   model of self-similar, self-gravitating, collapsing due to
   radiative viscosity yields a collapse timescale,
   $\tau_\gamma$, and suggests that the final phase of a Newtonian
   radiation dominated collapse to a black hole is slow. This example
   emphasizes the importance of other viscosity mechanisms like
   magnetic fields before destabilizing  GR effects take over; a
   detailed model is currently under study.

The above model can be easily adapted to the case of collapse due to
radiation drag at high redshift; the corresponding angular momentum
equation can be written as
\begin{equation}
\Sigma \del_t v_\phi +\Sigma (v_r/r) \del_r(r v_\phi)= -\kappa \Sigma
v_\phi,
\end{equation}
where $\kappa(t)= {4 \over 3} \epsilon_\gamma(t) \sigma_T/(c m_p)$ is
the coefficient of Compton drag in a completely ionized plasma and
$\epsilon_\gamma(t)= a T^4(t)$ is the CMBR energy  density. We obtain the
solution $\chi(t)= \exp{(\int \kappa \diff t/3)}$, and hence
$\chi_1(t)=\exp{(-(2/3)\int \kappa \diff t)}$; further using $\dot{z}=
-H_0 (1+z)^{5/2}$ appropriate for a matter dominated era, the collapse
factor due to Compton drag is given by
\begin{equation}
1/\chi_1(z)=  \exp{\left(\frac{16}{45} H_0^{-1}\frac{a}{m_p c} \sigma_T
T_0^4 \left [ (1+z_i)^{5/2}-(1+z_f)^{5/2} \right  ] \right)} =
\exp{\left (\frac{2\kappa(z_i)t(z_i)}{5}  \left [ 1-\left
(\frac{1+z_f}{1+z_i} \right)^{5/2} \right  ]\right)}
\end{equation}
which shows that the e-folding time in the angular momentum at $z_i
\simeq 1400$,  is initially  shorter than the Hubble time by two
orders of magnitude- $\kappa(z_i)t(z_i)=2.2 \times 10^2  \left
((1+z_i)/1400\right)^{5/2}$, a result that is consistent with
Loeb(1994) who computed a spherical model of a cloud in an expanding
background (here it is assumed that the disk has turned around and
collapsed). Therefore CMBR drag is  effective at redshifts
$1400>z\gsim 300$ and it is possible to collapse smaller mass  clouds
$\lsim 10^5 \Msun$.
\section{Supermassive stars}
\label{stars}
Some of the proposed scenarios are envisioned to lead to a build up of
a supermassive star at relativistically compact scales as an
intermediate  stage of the evolution before the gravitational
instability sets in and  a rapid final collapse to a supermassive
black hole ensues.

Supermassive stars (SMS) are equilibrium configurations that are
dominated by radiation pressure (the luminosities are nearly at the
Eddington limit) and can have masses  between $10^4 \Msun$ and about
$10^8 \Msun$. They are  expected to be fully convective (Loeb \& Rasio
(1994) give a formal argument that radial entropy has to develop
eventually which drives convection), isentropic and their structure
can be well described by a Newtonian polytrope with $\gamma=4/3$. The
energy per nucleon in an SMS is given by the radiation entropy in
units of the Boltzmann constant ($S/k \sim  0.94 (M/\Msun)^{1/2}$
where $M$ is the mass of the star (see eg. Zeldovich \& Novikov 1967).
The evolution proceeds on a Kelvin-Helmholtz timescale and is driven
by loss of energy and in the case of rotating SMSs, loss of angular
momentum through mass-shedding.  Pressure contributions from plasma
components raise the adiabatic index of the equation of state, $\Gamma
= 4/3 + \beta/6$ marginally above the critical value 4/3 where $\beta$
is ratio of gas to radiation pressure. General relativity leads to the
existence of a maximum for the equilibrium mass (as a function of the
density) and  gravitation instability sets in when $\Gamma$ falls
below a critical value $\Gamma <\Gamma_c  \approx (2/3)
(2-5\eta)/(1-2\eta)+ 1.12 R_s/R$ where $\eta=T/|W|$ is the ratio of
the  rotational to the  gravitational potential energy and $R$ is the
radius of the star (Misner, Thorne \& Wheeler 1973).  If the plasma
contribution is not enough, then the star shrinks during the
evolution; rotation can however hold up the collapse if $\eta$ is
above a critical value. The SMS evolves on a Kelvin-Helmholtz
timescale $t_{KH} =|E_c|/L_{edd}\simeq 10^9/(M/\Msun)$yrs where $E_c=3
\times 10^{54}$ergs is the equilibrium energy at the onset of
gravitational instability.  Typical SMSs with $M\sim 10^6 \Msun$ have
a lifetime of a 1000 yrs. Rotation can appreciably stretch their
equilibrium evolution; Baumgarte and Shapiro (1999) found a lifetime
independent of stellar mass of ($9 \times 10^{11}$s) and that key
non-dimensional ratios, $R/R_s, \eta,$ and $Jc/(GM^2)$ for maximally
and rigidly rotating polytropes are independent of the mass, spin or
radius of the star. Driven by radiation and angular momentum loss
through a mass-shedding sequence, the SMSs collapse leads to a
formation of a black hole through an explosion powered by hydrogen
burning in CNO cycle and associated with gigantic release of
neutrinos. However, this neutrino  flux and resulting background from
such sources is weak for even the new generation neutrino detectors
like Super Kamiokande; but the possibility that about 10\% of the
baryons are locked in SMSs  at $z<1$ (Shi \& Fuller 1998) can
potentially be ruled out.

\section{Summary and discussion of observational discriminants}
Based on the arguments here models of monolithic formation and
collapse (Mangalam 2001, 2003) of supermassive disks can explain the
mass and redshift distribution of black holes of $z \lsim 6$. However
plausible merger scenarios (with dynamical friction) of smaller
$\lsim 100 \Msun$ objects that form around $z\simeq 20$, probably
through the direct collapse and supermassive star route are needed for
highest redshift quasars.   In the case of self-similar
self-gravitating contraction due to radiation viscosity, the timescale
turned out to be a fraction of Hubble time, $\tau_\gamma \sim{\sigma_T
c/(m_p G)}  = 6 \times 10^9$yrs; clearly, one needs to take into
account GR effects which is a future goal. The case of Compton drag
yielded a timescale that is two orders of magnitude smaller than the
Hubble time and is effective in the range $1400>z\gsim 300$.  Hence,
it is possible to collapse smaller mass clouds $\lsim 10^5 \Msun$. It
worth investigating the statistics of such collapsed objects and
whether mergers can produce the high redshift quasars in the picture
above.  The luminosity function needs to be more precisely determined
to help distinguish between the models and the resulting stellar cusp
profiles of the merged systems are likely to  be different. The
observed relation, $\bh\propto\sigma^\alpha$, with $\alpha= 4-5$ can
be explained on the basis of the following physical arguments of
saturation of black hole mass in (Silk \& Rees 1988, Wyithe \& Loeb
2002): the black hole mass saturates when luminosity impedes further
accretion; ie, $L_{edd} \propto B. E./{t_{dyn}}$, where the
gravitational binding energy scales as $M^2/R$ resulting in
$\alpha=5$. Alternatively, though in a similar vein,  the fraction of
stars in an isothermal distribution that is captured by a
Schwarszchild black hole  is given by $f(r) \simeq (J_{\rm cap}/(2
\sigma r))^2$, where $J_{\rm cap}=4 GM/c$ is the maximum  angular
momentum for capture; this translates into a energy flux (evaluated
near the radius of  influence, $r_h$) $\propto \rho(r_h) r_h^2  \sigma
f(r_h)$ which again results in $\alpha=5$ (Zhao, Haehnelt \& Rees
2002).

In models of disk contraction that depend on self-gravity induced
instabilities, the accretion is effective only upto the point when the
Keplerian potential dominates over the gravity of the disk, which
implies final black hole masses $\bh > 10^5 \Msun M_{d9}$ where
$M_{d9}$  is the disk mass in units of $10^9 \Msun$ (Mangalam 2001,
Loeb \& Rasio 1994). This has direct  implications for the seed
mass. In conclusion, our understanding of the process is only
beginning and there several promising ideas that need to be further
explored and more observations tests are required for discriminating
amongst the models.

\label{lastpage}
\end{document}